\documentclass[12pt,a4]{article}
\usepackage{graphicx}
\usepackage{epstopdf}
\usepackage{wasysym}

\title{ Feasibility Studies of Exclusive Diffractive Bremsstrahlung Measurement at RHIC Energies
\author{
Janusz Chwastowski$^1$, Antoni Cyz$^2$, \\{\L}ukasz Fulek$^3$, Rados{\l}aw Kycia$^1$,  
\\Bogdan Pawlik$^2$\footnote{Corresponding author, e-mail address:Bogdan.Pawlik@ifj.edu.pl}, Rafa{\l} Sikora$^3$, Jacek Turnau$^2$\\[8pt]
$^1$Dept. of Physics, Mathematics and Computer Science,\\
Cracow University of Technology,
Warszawska 24,\\
31-155 Krak\'ow, Poland\\[8pt]
$^2$The Henryk Niewodnicza\'nski Institute of Nuclear Physics,\\
Radzikowskiego 152,
31-342 Krak\'ow, Poland\\[8pt]
$^3$Faculty of Physics and Applied Computer Science AGH--UST,\\
al. A. Mickiewicza 30,
30-059 Krak\'ow, Poland
}
}
\begin{document}
\maketitle
\abstract{
Feasibility studies of an observation of the exclusive diffractive bremsstrahlung at RHIC  at $\sqrt{s} = 200$~GeV and 
at $\sqrt{s} = 500$~GeV are reported. A simplified approach to the photon and the scattered proton energy 
reconstruction is used. Influence of possible backgrounds is discussed.
}
  
\section{Introduction}
Electromagnetic bremsstrahlung is widely used in various applications. In high energy physics it became a very attractive tool 
with advent of HERA. Owing to its simple and easy to register final state and relatively large and precisely calculable cross-section
it served as an efficient tool for the determination of absolute and instantaneous luminosities of the machine as well as an efficient
beam diagnostic and monitoring tool  (see for example \cite{zeus}). The measurements were based on the angular properties 
of the radiated photons.

In 2011, V. Khoze  \textit{et al.} \cite{khoze1} proposed to supplement the LHC forward physics programme with the detection
of elastic scattering of protons accompanied by radiative photons, the diffractive bremsstrahlung. One should note that diffractive 
bremsstrahlung was never studied at accelerator energies. Later, phenomenological investigations of exclusive diffractive photon 
bremsstrahlung in proton-proton interactions at large energies were considerably extended by Lebiedowicz and Szczurek \cite{szczurek}.
The most important extension of \cite{khoze1} was the inclusion of effects of the proton finite size, \textit{i.e.}, the formfactor. In their 
model values of the parameters were based on an educated guess and are a subject to experimental verification.
In fact, the Donnachie--Landshoff parameterisation \cite{dl} was used with linear pomeron trajectory with intercept of 1.0808 and 
the slope of 0.25 GeV$^{-2}$. The elastic slope energy evolution was described as 
$$
B(s) = B^{NN}_{IP}+2\cdot\alpha^\prime_{IP} \ln{\frac{s}{s_0}}
$$
with $s_0 = 1$~GeV$^2$ and $ B^{NN}_{IP} = 9$~GeV$^{-2}$. Also, the cut-off parameter of a form factor related to the off-shell 
effects was set to 1~GeV. One should notice that this form factor plays important role for the invariant mass of the $\gamma~p$ system 
above 1~GeV which is not important for the present considerations. For the full account of the model parameters see \cite{szczurek}.
\\
Lebiedowicz and Szczurek
calculated also other processes leading to the exclusive $pp\gamma$ final states, \textit{e.g} virtual photon re-scattering, however they found
that these processes do not play an important role in the extremely forward direction investigated in the present study. The measurements
of exclusive diffractive bremsstrahlung can be considered as complementary to the luminometers and luminosity monitors proposed 
in~\cite{khoze2, krasny1}.

The exchanged Pomeron ensures the energy-momentum conservation in the diffractive bremsstrahlung process:

$$p+p \rightarrow p+p+\gamma.$$

 The cross-section is quite large and of the order of micro-barns. The photon angular distribution resembles the one observed for 
 the classical, electromagnetic bremsstrahlung 
 
 $$ \frac{d\sigma}{d\Theta_\gamma} \sim \frac{\Theta_\gamma} {\left(\frac{m_p^2}{E_p^2}+\Theta_\gamma^2\right)^2},$$
 \noindent
 where $\Theta_\gamma$ is the polar angle of the emitted photon, $m_p$ is the proton mass and $E_p$ its energy.
 The photon angular distribution is peaked in the forward direction with a characteristic unit of  $\sim 1/\gamma$ ($\gamma$ 
 is the radiating particle Lorentz factor). One should also note that the scattered proton angular distribution is 
 extremely narrow due to the large value of the nuclear slope parameter at high 
 energies. Both final state particles can be registered in the dedicated parts of the detector located at large rapidities.

In this paper a feasibility study of diffractive photon bremsstrahlung measurement at RHIC energies is carried out assuming the STAR
detector Phase II configuration. Registration of bremsstrahlung photons in the Zero Degree Calorimeter (ZDC) \cite{zdc} and the scattered
protons in the STAR Roman Pots \cite{rp} is considered. Simulation of the considered process was performed using a dedicated generator 
\cite{kycia2} which extends an earlier one, GenEx,  described in \cite{kycia1}. This generator is based on calculations presented in
\cite{szczurek}. Simulation of the experimental apparatus is simplified and only its basic properties, such resolutions, are used.
 
\section{Experimental Set-up}

As was already mentioned the most important parts of the apparatus for present study are the Zero Degree 
Calorimeters ~\cite{zdc} and the Roman Pot stations \cite{rp}. 

Two ZDCs are placed symmetrically with respect to the nominal interaction point (IP) at the distances of 1800 cm. They were 
designed to detect and to measure the total energy of neutral particles emitted within a small solid angle in extremely forward 
directions. The ZDC extends from -5~cm to 5~cm in the horizontal direction and from -5~cm to 7.5~cm in the vertical one.
The ZDCs are the Cherenkov-light sampling calorimeters. They consist of three modules and are approximately 5 interactions
lengths deep. They can serve as triggering devices.  The ZDCs were upgraded with the Shower Maximum Detectors (SMD)
located between the first and second module. The SMDs delivers the hadron shower position in the plane perpendicular to the beam  axis.

There are four stations of the Roman Pots placed symmetrically with respect to the IP for the STAR Phase II programme. The stations are
located between the RHIC DX and D0 magnets. The DX is a dipol magnet and helps to analyse the scattered proton momentum. Each
station contains 10 planes of silicon strip detectors with alternating direction of the strips inserted horizontally, approximately in the RHIC 
ring plane. The spatial resolution of the space-point measurement is about 30~$\mu$m. The distance between the detector active part and
the beam plays a crucial role in the measurement and decides about the detector acceptance. In real running conditions this distance is a
compromise between the accelerator and detector safety and the beam related backgrounds.

\section{Final State Properties}
Samples of diffractive bremsstrahlung events were generated for the two centre of mass energies used to measure proton-proton 
interactions  at RHIC, namely 200~GeV and 500~GeV. Each sample contains 1 000 000 events. In the generation it was requested 
that the photon is emitted in the $+z$ direction. The radiated photon energy varied between $0.5$~GeV and $20$~GeV at 
$\sqrt{s} = 200$~GeV, and $0.5$~GeV and $50$~GeV at $\sqrt{s} = 500$~GeV. Table \ref{tab1} lists centre of mass energies, inverse of
the Lorentz factor, MC predicted cross-section and the average polar angle of the emitted photon characterising the generated samples. 
\begin{table}[ht]
\begin{center}
\caption{Quantities characterising the generated Monte Carlo samples.}
\label{tab1}
\vspace{0.5cm}
{\centering
\begin{tabular}{|c|c|c|c|}
\hline
 $\sqrt{s}$ [GeV]&  $1/\gamma$ &\multicolumn{2}{|c|}{Monte Carlo}\\
 \cline{3-4}
 & &  $\sigma_{MC}$ [$\mu$b] & $<\Theta_{\gamma}>$\\
\hline
200 &  0.00938& 0.6007$\pm$0.0004&0.011\\
\hline
500 & 0.00375& 0.9770$\pm$0.0007 & 0.0045\\
\hline
\end{tabular}
}
\end{center}
\end{table}
The average emission angles of the photon are in a good agreement with the inverse of the Lorentz factor of the colliding proton.

\begin{figure}[htb]
\centerline{
\includegraphics[width=58mm,height=58mm]{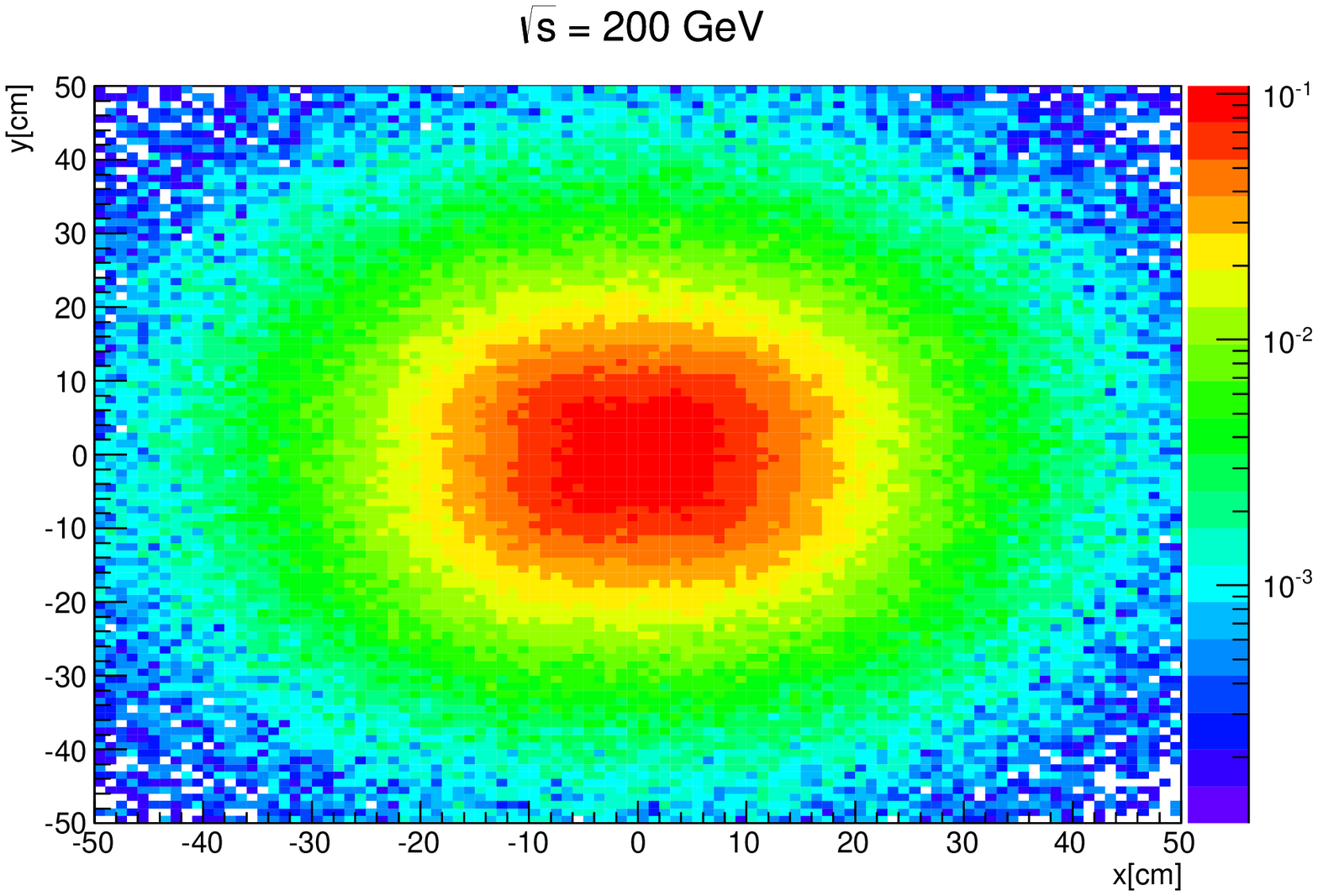}
\includegraphics[width=58mm,height=58mm]{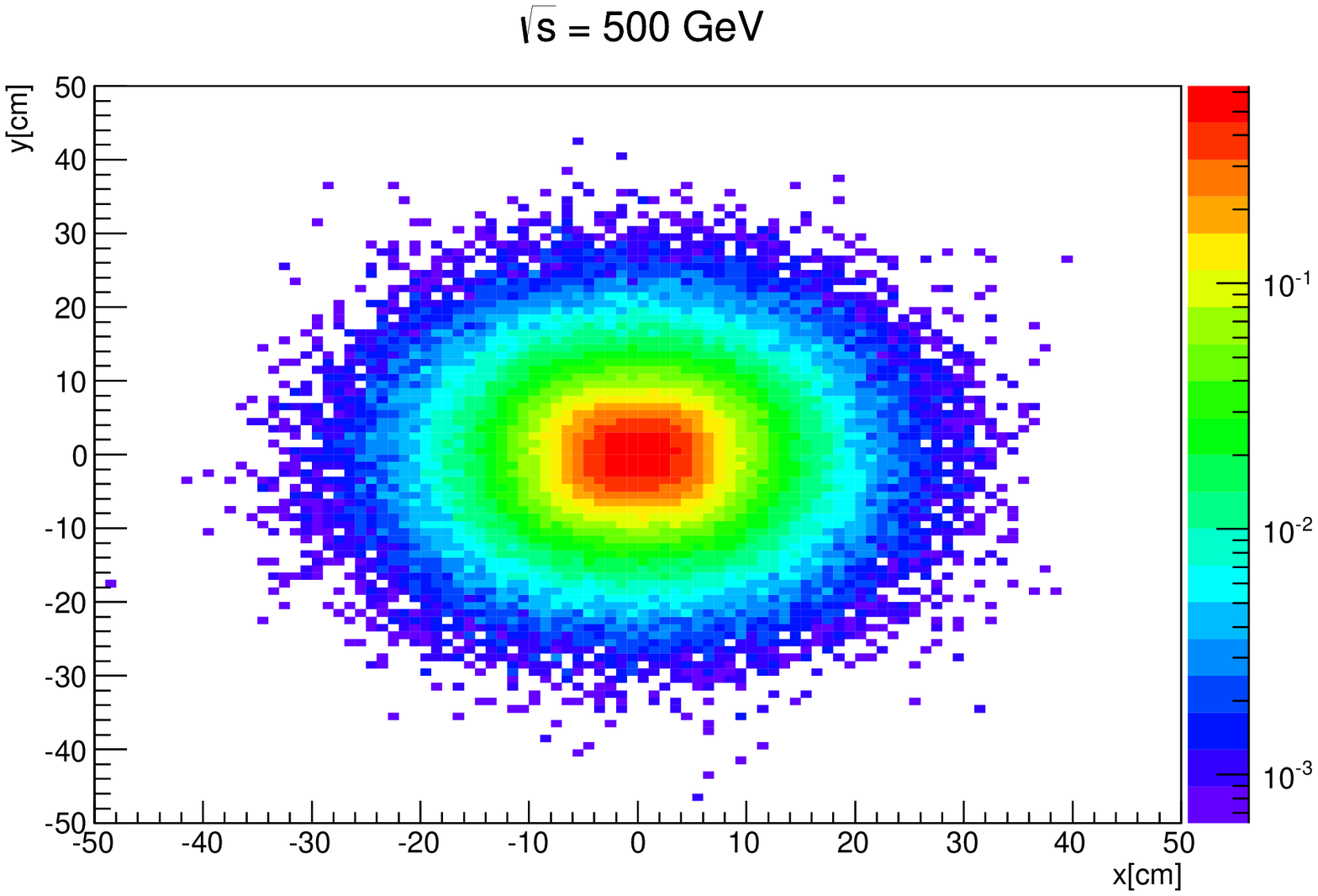}
}
\caption{Distribution of the photon impact point position at the ZDC face for $\sqrt{s} = 200$~GeV (left) and $\sqrt{s} = 500$~GeV
 (right). Color scale is in percent. }
\label{fig:GamYX}
\end{figure}

The distribution of the diffractive bremsstrahlung photon position in the plane transverse to the collision axis at the ZDC 
location is shown in Figure \ref{fig:GamYX} for 
proton-proton interactions at $\sqrt{s} = 200$~GeV. 
One can observe a characteristic picture. Majority of the photons hits the ZDC plane within the circle with a radius of about 
25~cm. As was already mentioned the ZDC geometric acceptance allows the registration of only a part of these photons.
At $\sqrt{s} = 500$~GeV the percentage of photons reaching the ZDC increases considerably as the radiated photon angular 
distribution peaks more in the forward direction.

\noindent
The relative energy loss
$$ \xi = \frac{E_p-E_p^\prime}{E_p} $$
is a handy variable describing the scattered proton. 

\begin{figure}[htb]
\centerline{
\includegraphics[width=58mm,height=58mm]{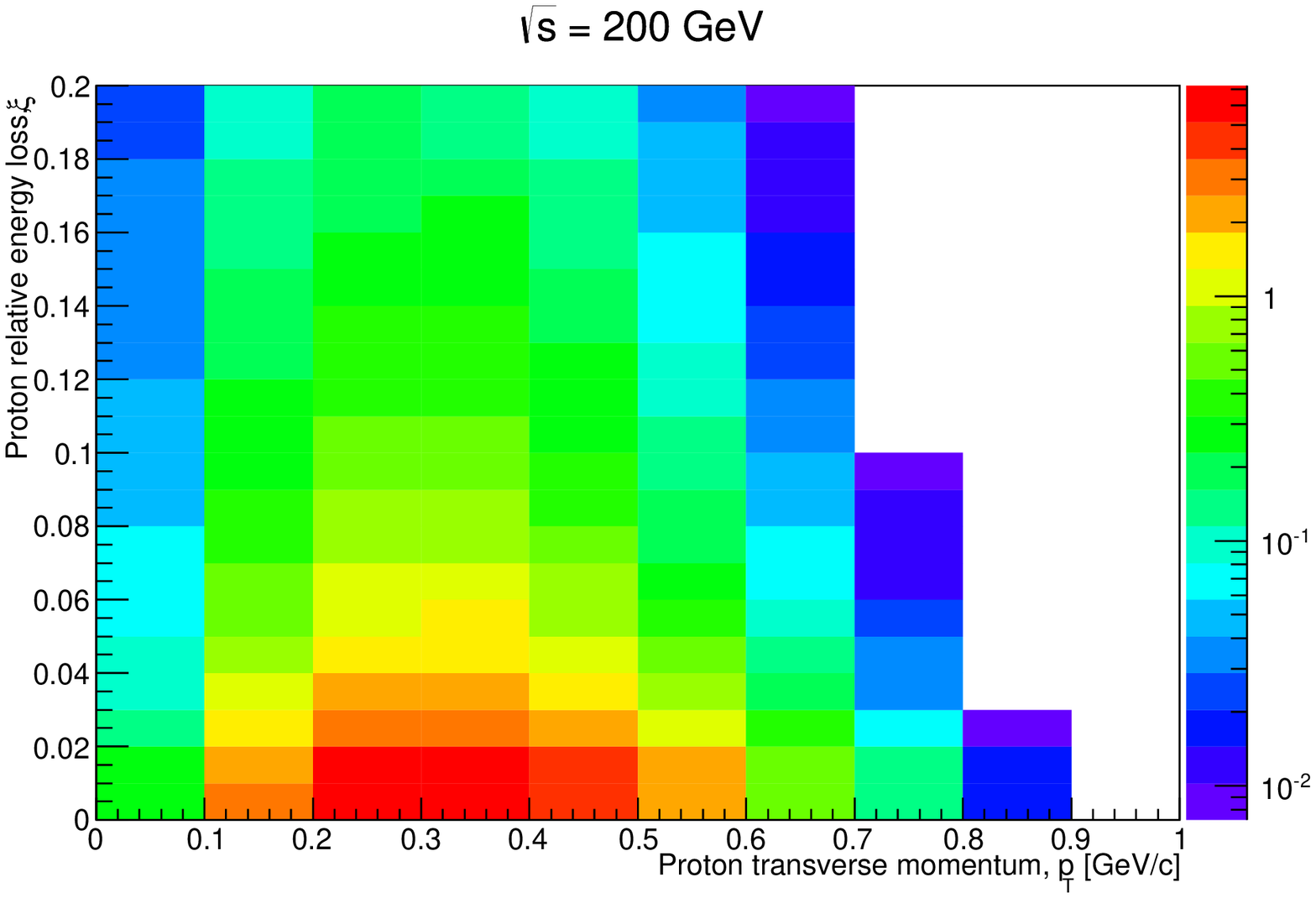}
\includegraphics[width=58mm,height=58mm]{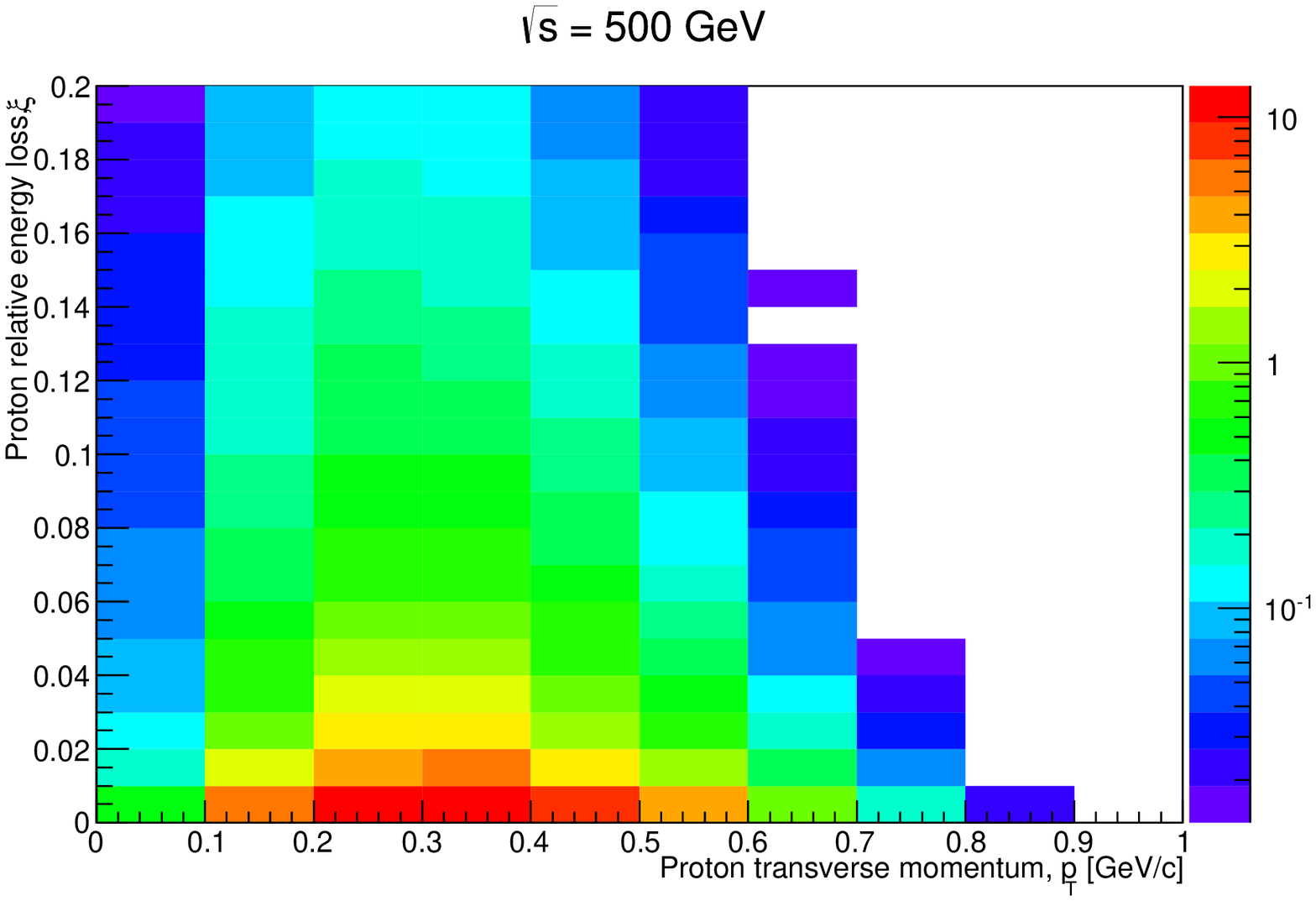}
}
\caption{Correlation between the relative energy loss of a proton, $\xi$, and its transverse momentum, $p_T$ for $\sqrt{s} = 
200$~GeV (left) and $\sqrt{s} = 500$~GeV (right). Color scale is in percent.}
\label{fig:KsiPt}
\end{figure}

The correlation plots of the scattered proton transverse momentum, $p_T$, versus its  relative energy loss  in diffractive bremsstrahlung
events at the centre of mass energies of 200~GeV and 500~GeV are presented in  Figure~\ref{fig:KsiPt}.
As can be observed a vast majority of the scattered protons is contained within the region limited by 
$ 0.1$~GeV/c $ < p_T <0.6$~GeV/c and $\xi < 0.1$. However, one should notice that  some part of protons with transverse 
momentum  within the interval $(0.2; 0.4)$~GeV/c is  characterised by the relative energy loss reaching up to 0.2.
A similar picture is seen for the centre of mass energy of 500 GeV. Apparent peaking of the distribution at small $\xi$ values 
reflects its minimum accessible value -- fixed lower limit on the radiated photon energy. Higher, up to 0.2, $\xi$ values are observed for 
$p_T \in (0.1;0.5)$~GeV/c.

\section{Analysis}

Present study was performed using the above mentioned Monte Carlo generated samples. 
In the calculations the final state particles were transported using an application based on the Geant4 \cite{geant4} code.  This
application makes use of the MAD-X \cite{mad} description of the RHIC magnetic lattice. The scattered protons were
transported checking whether they will reach the silicon detectors placed in the Roman  Pots. A simplified  simulation of the
detector response was used to create the hits and a simplified reconstruction of the scattered proton trajectory was performed. The
 reconstruction efficiency of a proton candidate track is about 98\% \cite{sikora}. 

The signal signature is an energy deposit in the ZDC accompanied by the Roman Pot track candidate in the same ($+z$) hemisphere 
of the  reaction. The sum of the reconstructed  energies of a proton and that of the photon should be close to the incident proton 
energy. Also, the STAR detector should not register an interaction.
 
One should note that the ZDC is dedicated to the measurement of neutral hadrons, mainly neutrons, to trigger the 
ultraperipheral ion-ion interactions.  Therefore, its quality of the electromagnetic measurements is rather limited. For the 
purpose of present analysis it was assumed that  the energy measurement resolution is 30\%/$\sqrt{E}$ for photons with 
energy above 1~GeV and that the photons with smaller energies do not trigger the ZDC readout at all.
 If one demands in addition a realistic distribution of the interaction vertex and that the reconstructed photon energy is 
 above 1~GeV then the  fraction of accepted events is about 18\% and  76\% at the centre of mass energy of 200~GeV and 
 500~GeV, respectively.
 
 The measurement conditions concerning the scattered proton have a large impact on the visible cross-section. Here, a crucial 
 role is played  by the distance between the silicon detector edge and the beam position at the detector location. It is clear that 
this distance defines the minimum value of the measurable relative energy loss of a proton. Table \ref{distance} lists the 
fraction of diffractive bremsstrahlung events having the ZDC energy above 1~GeV and an associated  proton track
 reconstructed using  the Roman Pots’ measurements. The statistical errors on the quoted numbers are negligible.
 \begin{table}[ht]
\begin{center}
\caption{Fraction of accepted events with  the ZDC energy above 1~GeV and an associated track in the Roman Pot 
as a function of the silicon detector-beam distance. }
\label{distance}
\vspace{0.3cm}
{\centering
\begin{tabular}{|c|c|c|}
\hline
distance &\multicolumn{2}{|c|}{ fraction of events [\%]}\\
\cline{2-3}
[mm]   &  $\sqrt{s} = 200$~GeV  &$\sqrt{s} = 500$~GeV\\
\hline
15&  $3.5$ & $19.1$\\
\hline
20& $3.2$& $10.9$ \\
\hline
25& $3.0$& $5.3$ \\
\hline
\end{tabular}
}
\end{center}
\end{table}

In the experimental procedure of an event selection the energy conservation relation should be formulated as:
 \begin{equation}
 |E_{beam} - E_{\gamma, ZDC}-E_{p, RP}^\prime| < \delta_{r},
\label{eq:Econs2}
  \end{equation}
where $E_{\gamma, ZDC}$ is the photon energy seen by the ZDC, $E_{p, RP}^\prime$ is the reconstructed proton energy and
$\delta_{r}$ is the accepted width of the energy conservation requirement. The value of $\delta_{r}$ reflects both the photon 
and the scattered proton energy reconstruction resolutions. It was checked that in the generation the \textit{non-radiating} proton energy 
does not deviate  from the beam energy by more than 0.1\permil~~in the considered kinematic domain and hence this factor can be safely neglected in Eq. (\ref{eq:Econs2}).

\begin{figure}[ht]
\centerline{
\includegraphics[width=58mm,height=58mm]{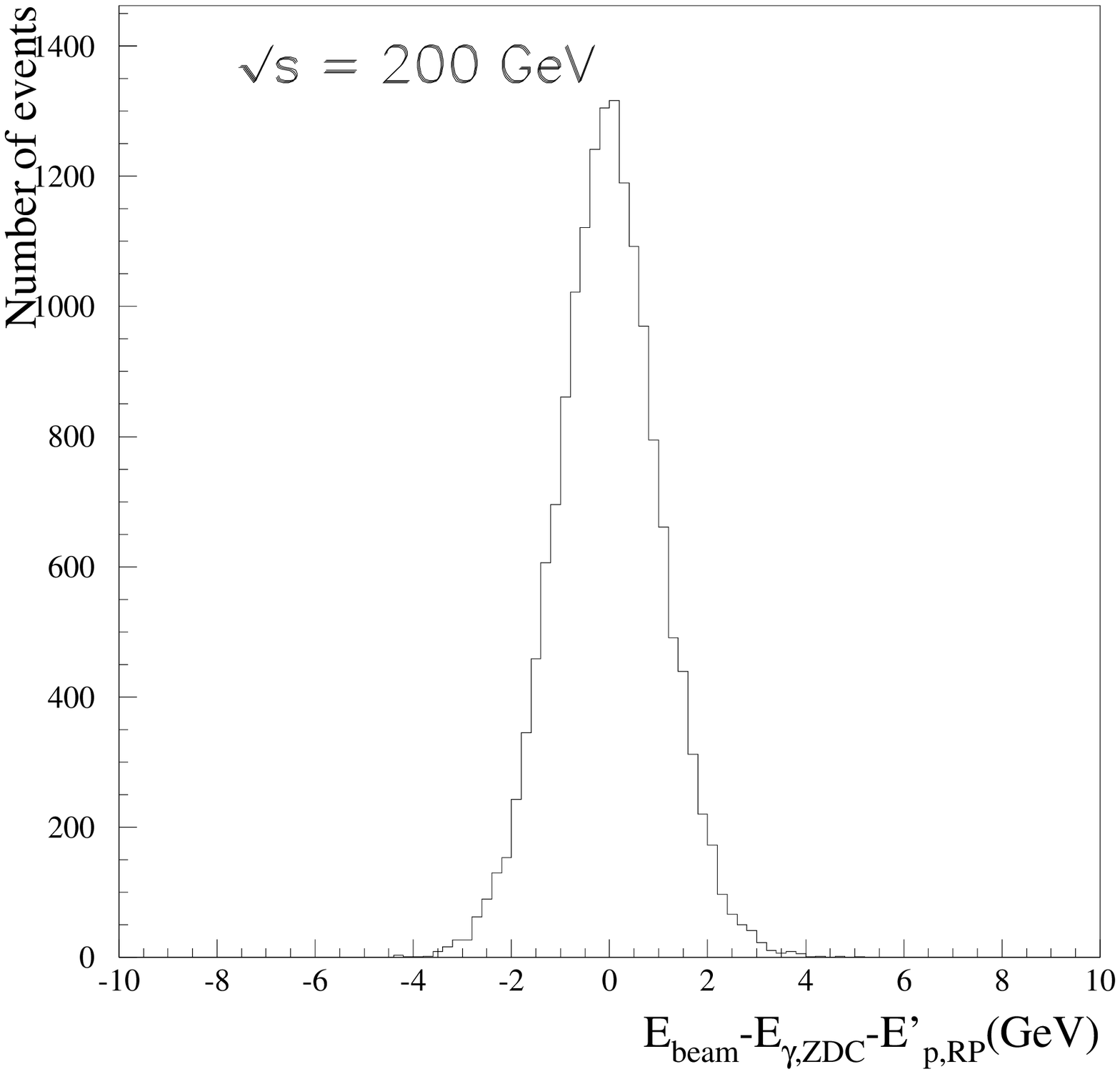}
\includegraphics[width=58mm,height=58mm]{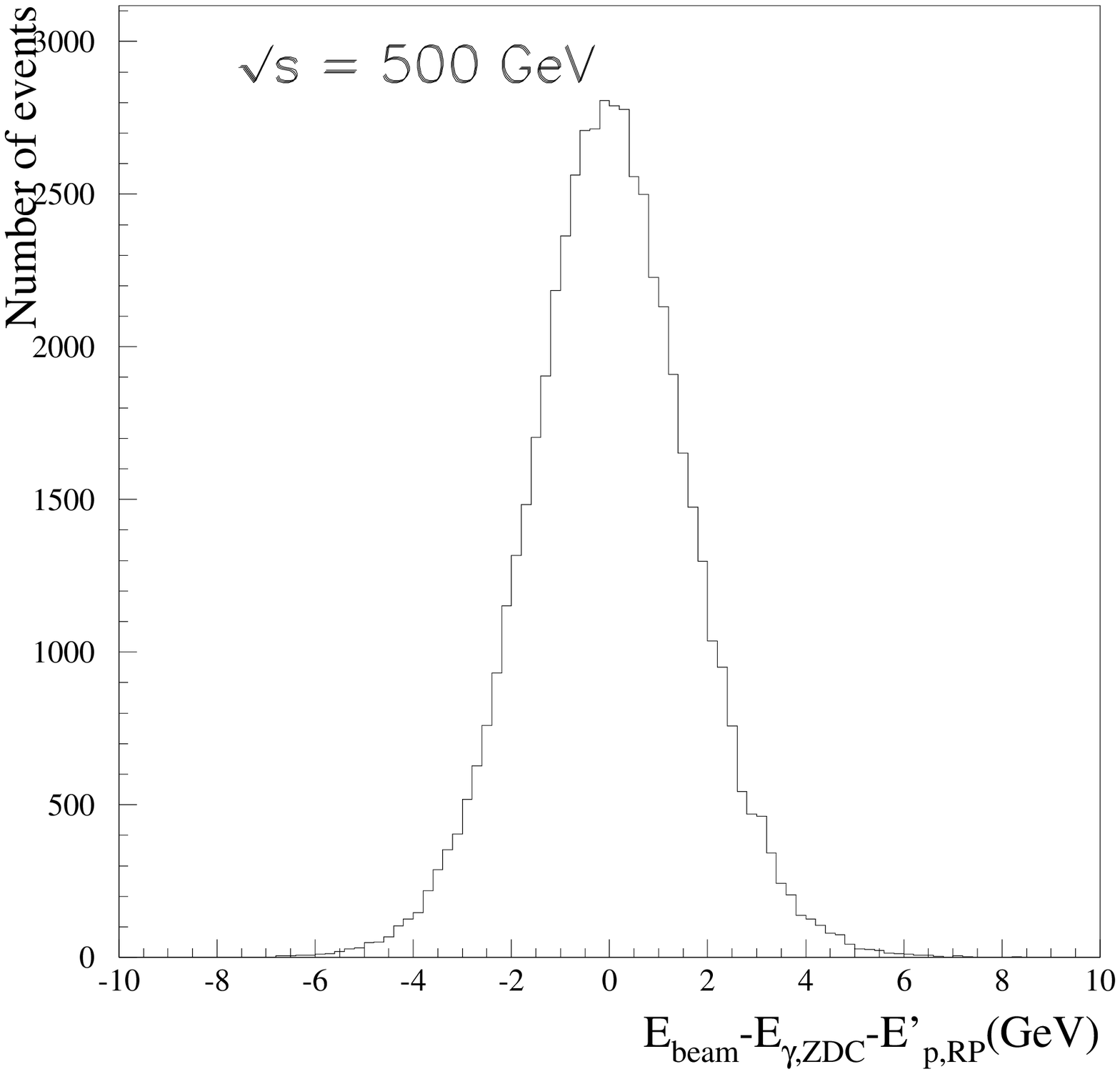}
}
\caption{Distribution of the energy conservation, $E_p - E_{\gamma, ZDC}-E_{p, RP}^\prime $ for the silicon detector-beam 
distance of 20~mm for the centre of mass energy of  200~GeV and 500~GeV, and the proton energy reconstruction resolution 
of form of $8\%\cdot\sqrt{E}$. }
\label{fig:Econs}
\end{figure}

The distribution of $E_{beam} - E_{\gamma, ZDC}-E_{p, RP}^\prime$ is presented in Figure~\ref{fig:Econs}
 for events having $E_{\gamma, ZDC} >1$~GeV and assuming the silicon detector-beam distance of 20~mm for both considered values 
 of the centre of mass energy.
In the calculations the ZDC energy was reconstructed as described above and the proton energy reconstruction resolution of the form of $8\%\cdot\sqrt{E}$ was assumed which delivers uncertainties higher than observed earlier ~\cite{rafal}. The RMS values of the distributions shown in Figure~\ref{fig:Econs} are 1.07~GeV and 1.65~GeV at 
$\sqrt{s} =   200$~GeV and   $\sqrt{s} =   500$~GeV, respectively.  Eventually, value of the $\delta_{r}$ parameter was set to 
the triple of the corresponding value. Table \ref{distance1} lists the fractions of the accepted, \textit{fully reconstructed}  events as a
function of the beam-detector distance.
  \begin{table}[ht]
\begin{center}
\caption{Fraction of events passing the energy conservation constraint --  Eq. (\ref{eq:Econs2}) -- as a function of the 
silicon detector-beam distance assuming $8\%\cdot\sqrt{E}$ resolution of the proton energy reconstruction.}
\label{distance1}
\vspace{0.5cm}
{\centering
\begin{tabular}{|c|c|c|}
\hline
distance & {$\sqrt{s} = 200$~GeV}     &  {$\sqrt{s} = 500$~GeV} \\  \cline{2-3}
 [mm]    & $\delta_{r} = 3.21$~GeV    & $\delta_{r} = 4.95$~GeV \\
\hline
15&  1.7\% & 9.5\% \\
\hline
20& 1.6\%&5.4\% \\
\hline
25& 1.5\%& 2.6\% \\
\hline
\end{tabular}
}

\end{center}
\end{table}

The beam related pile-up brings in limitations to the measurement. It is clear that an ideal occurrence of diffractive bremsstrahlung
would be in a bunch crossing in which there is no additional strong force mediated proton-proton interactions. The probability that there 
are no additional strong force mediated interactions  
depends on the total $pp$ inelastic cross-section and the instantaneous luminosity delivered be the machine. 

\begin{figure}[htb]
\centerline{
\includegraphics[width=58mm,height=58mm]{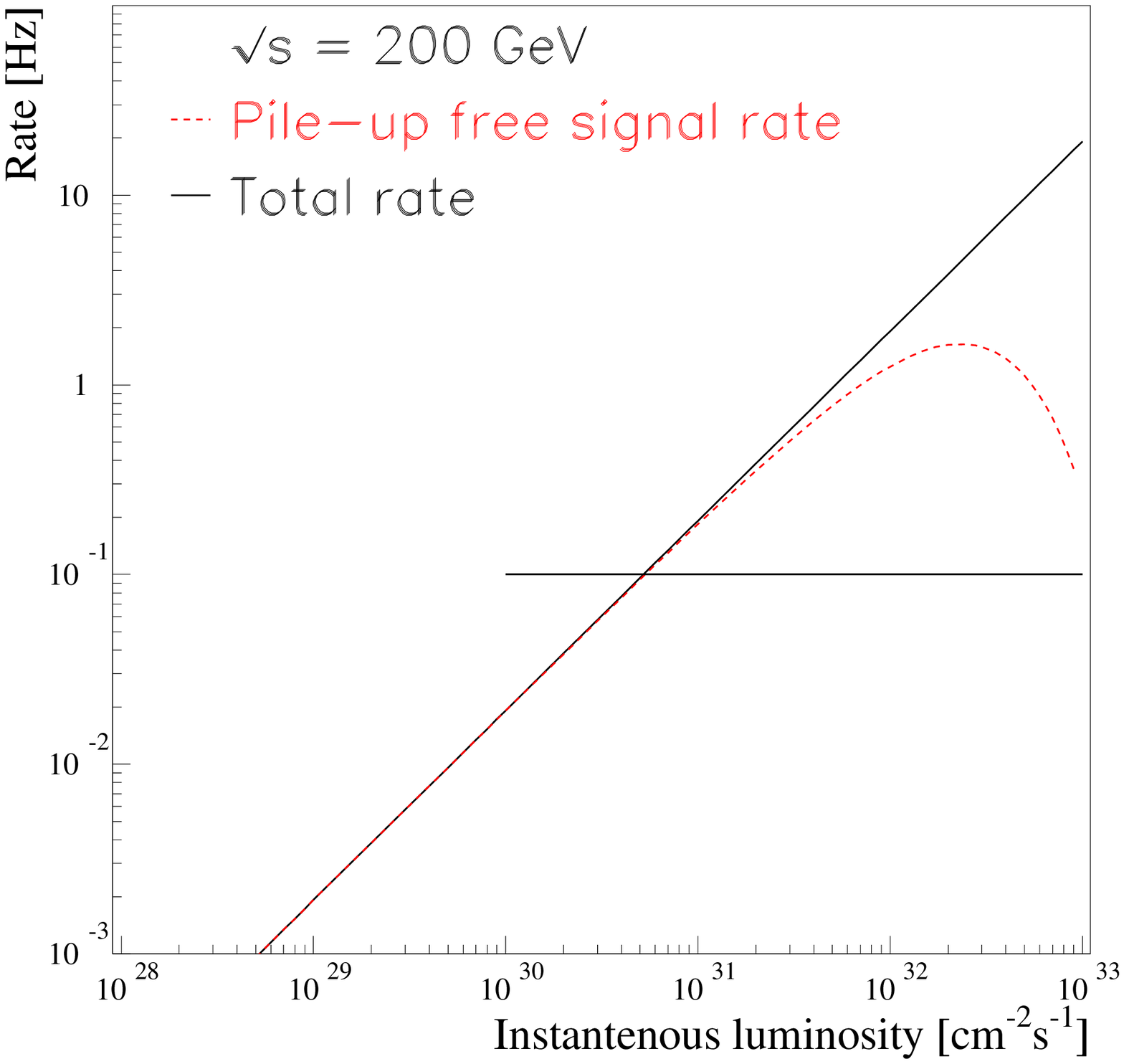}
\includegraphics[width=58mm,height=58mm]{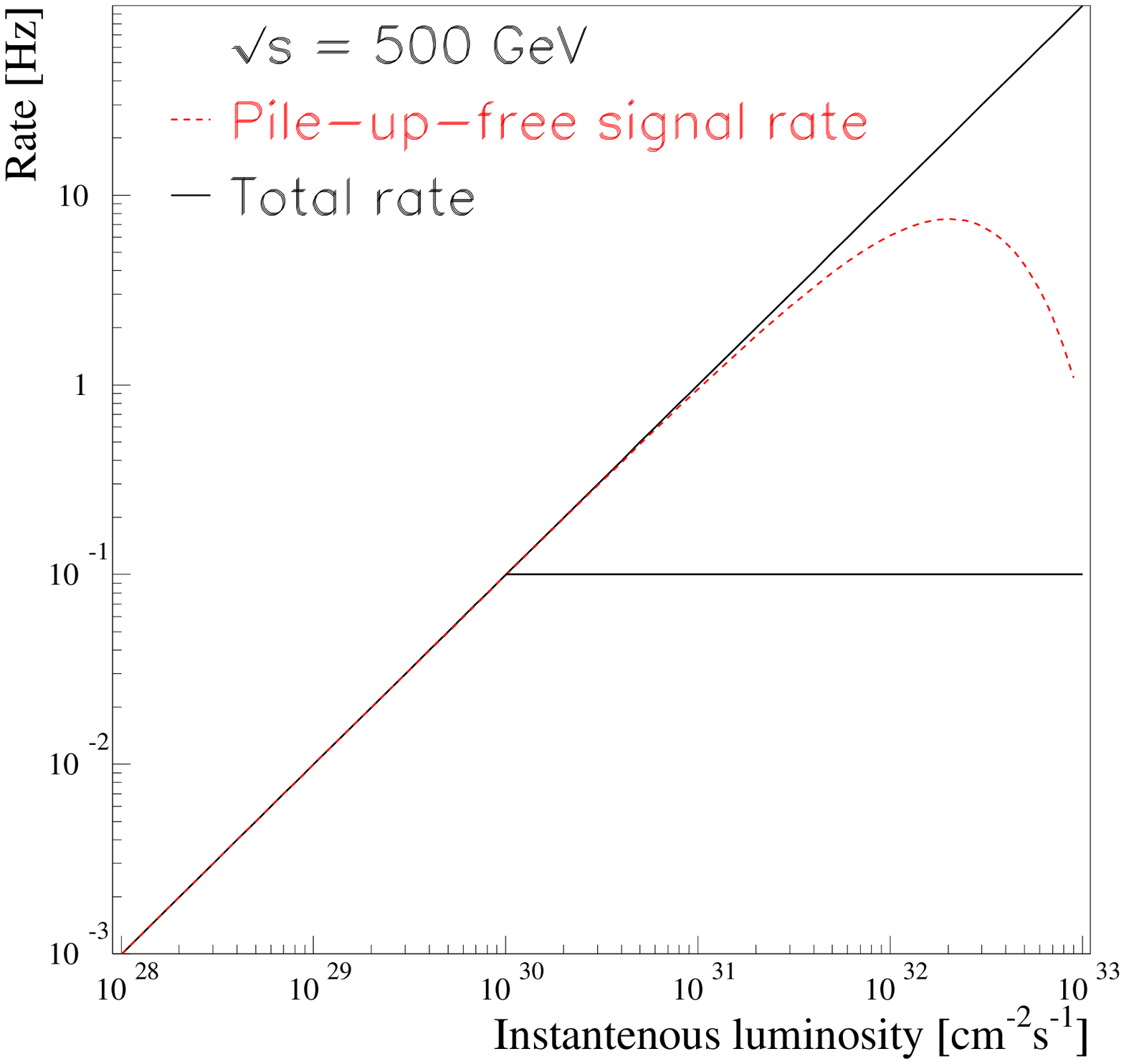}
}
\caption{The rate of the diffractive bremsstrahlung events as a function of the machine instantaneous luminosity. 
The solid line  -- the total rate, the broken line -- the rate in pile-up-free bunch crossings. See text for details.}
\label{fig:rates}
\end{figure}

Comparison of the total and the pile-up-free diffractive bremsstrahlung event rates is shown in  Figure \ref{fig:rates} for the
 two considered values of the centre of mass energy. To calculate the predictions the silicon detector-beam distance was set to
 20~mm. In the calculations the total inelastic $pp$ cross-section  of 43~mb at $\sqrt{s} = 200$~GeV (49~mb at $\sqrt{s} = 
 500$~GeV) foreseen by  Pythia 8 \cite{pythia} generator was used. The luminosity was assumed to be evenly distributed over all 
bunch crossings. The diffractive bremsstrahlung cross-section was reduced to the visible one using the fractions listed in 
Table~\ref{distance1}. This procedure yielded the visible cross-section values of about 19.2~nb and 105.5~nb at the centre of mass
energy of 200 GeV and 500 GeV, respectively.

As can be observed in Fig.~\ref{fig:rates} the requirement of the rate above  0.1~Hz can be achieved for the instantaneous
luminosity values ${\cal L} > 8\cdot 10^{30}$~cm$^{-2}$s$^{-1}$ for the centre of mass energy of 200 GeV. In case of $pp$
interactions at $\sqrt{s} = 500$~GeV the rate above 0.1~Hz is seen for ${\cal L} >  2\cdot 10^{30}$~cm$^{-2}$s$^{-1}$. Such a
rate implies a sample of at least 180 events collected in a typical, 30 minutes long STAR data acquisition run. The maximum rates are
reached for ${\cal L} \approx 2\cdot 10^{32}$~cm$^{-2}$s$^{-1}$ and are about 1~Hz and 4~Hz at the c.m. energy of 200~GeV
and 500~GeV, respectively. For larger values of instantaneous luminosity the accepted signal rates rapidly decrease what
illustrates the influence of the pile-up. 

One should note that classical, electromagnetic bremsstrahlung yields the final state of exactly the same signature as the  
diffractive one. Moreover, kinematic properties of the radiative photon and those of the scattered proton are nearly identical 
for both processes. However, in diffractive bremsstrahlung the distribution of the scattered proton transverse momentum is broader. 
The electromagnetic bremsstrahlung cross-section values calculated for $pp$ interactions at $\sqrt{s} = 200$~GeV and
$\sqrt{s} = 500$~GeV are, respectively, close to 12~nb and about 14~nb  in the kinematic domains considered in the 
present paper. These values are nearly 60 and 140 times smaller than the diffractive  cross-section values delivered by the 
diffractive bremsstranlung generator. Moreover, the visible cross-section for the process of  electromagnetic bremsstrahlung 
will be a subject to the same reduction factor as the diffractive one, \textit{i.e.}, the resulting values will be well below 1~nb.
Therefore, the influence of the electromagnetic bremsstrahlung was neglected in this work.

\section{Backgrounds}

The single diffractive events in which there is a neutral (electromagnetic) energy produced within the ZDC acceptance and a fast, 
forward proton created in the recombination process may have a signature of an energy deposit in the ZDC and an associated 
track in the Roman Pots and therefore imitate events of diffractive exclusive bremstrahlung. Study of this background type were 
performed using the PYTHIA 8 \cite{pythia} generated sample of 1 000 000 000 events.

The events were rejected from the analysis if they fulfilled the following criteria:
\begin{itemize}
\item presence of a charged particle with transverse momentum larger than 0.1~GeV/c and the pseudorapidity, $|\eta| < 1$ (the STAR TPC acceptance),
\item presence of charged particle in the pseudorapidity regions: $1.086 \leq |\eta| \leq 2$ (STAR end-cap calorimeter) or $3.3 \leq |\eta| \leq 5$ (STAR BBC counters),
\item presence of neutral particle with energy not smaller than 1 GeV in the above mentioned pseudorapidity regions.
\end{itemize}

 The remaining events were accepted for further analysis during which the following requirements were imposed:
 \begin{itemize}
 \item the total neutral electromagnetic energy reaching the ZDC was not smaller than 0.5 GeV,
 \item events contained a proton with energy greater than 60\% of the incident beam energy. 
 \end{itemize}
For events passing these criteria the energies of photons and neutral hadrons within the ZDC geometric  acceptance were calculated. 
In the subsequent analysis these energies were treated as if they were associated with a single photon or neutron and were assumed to
be measured with perfect resolution. One should note that such a procedure will tend to overestimate the background influence
delivering  the worst case scenario.Eventually, events were accepted as those imitating the diffractive bremsstrahlung
signature if they had the following properties:
 \begin{enumerate}
\item there was a proton with energy $E_p >0.8\cdot E_{beam}$ 
and the electromagnetic energy in the ZDC, $E_{EM,ZDC} > 1$~GeV in 
a given hemisphere and the neutral hadron energy seen by the ZDC in the same hemisphere is $E_{HAD,ZDC} < 1.0$~GeV,
\item in the hemisphere opposite to the signal one:  $E_{EM,ZDC} < 1.0$~GeV,  $E_{HAD,ZDC} < 1.0$~GeV and $E_p < 1$~GeV.
\end{enumerate}
Such requirements were fulfilled by small fractions of events: $584\cdot 10^{-9}$ at 200~GeV and $33 040\cdot 10^{-9}$ at 
500~GeV. If one requests that the proton hits the silicon detector in the Roman Pots and is reconstructed then this fraction drops to
$223\cdot 10^{-9}$ and $9 072\cdot 10^{-9}$, respectively. Requiring the energy conservation  
$|E_{beam}-E_{EM,ZDC}+E_p| < 3$~GeV implies further reduction to about $39\cdot 10^{-9}$ at $\sqrt{s} = 200$~GeV and
$330\cdot 10^{-9}$ $\sqrt{s} = 500$~GeV which correspond to the visible cross-section for background processes of about 2~nb
and 16~nb, respectively.

There is no doubt that the Zero Degree Calorimeter upgraded with an electromagnetic front part would help to achieve even better 
results. The effect would be twofold. On one hand a precise electromagnetic calorimeter would allow to lower the limit on the 
minimum photon energy. On the other hand it would also diminish the influence of the photon energy measurement on the 
$\delta_{r}$ parameter. Moreover, equipping the electromagnetic part with capability of a preceise shower position measurement would 
allow to discriminate the events with production of mesons decaying into multiple photons. Such events are mainly due to the production of 
$\pi^0$, $\eta$ and $N^*$. PYTHIA 8 delivers a good description of these processes based on the experimental data parameterisations. 
Moreover, owing to the relatively low energies of $\pi^0$, $\eta$ or $N^*$ the photons created in their decays are produced with quite large 
polar angles in the laboratory frame. 
Therefore, they have a considerable chance to miss the ZDC acceptance and in such a way an event would not pass the signal selection
criteria. 

\section{Signal/Background Ratio}
The expected signal to background ratio (S/B) was as a function of the instantaneous luminosity for both considered values of energy. 
In calculations of  the S/B ratio it is assumed that both the signal and the background events are created in pile-up free bunch crossings. 
The S/B values of 10 for $\sqrt{s} = 200$~GeV and nearly 100 for $\sqrt{s} = 500$~GeV  were obtained independently of the 
instantaneous luminosity. It has to be reminded that the useful range of the instantaneous luminosities is related to the minimal value of the
rate of signal events.

\section{Summary and Conclusions}

Feasibility study of the measurement of diffractive bremsstrahlung process in $pp$ interactions at the RHIC energies was carried out. 
This study shows the the visible cross-section would be of the order of 20~nb if the accepted proton energy is above 80\% of the beam 
energy and that of the photon is within ~$[1; 20]$~GeV interval at $\sqrt{s} = 200$~GeV. At the centre of mass energy of 500~GeV the 
visible-cross section is about 106~nb. These values imply the event rates above 0.1~Hz for the instantaneous luminosity of the
machine  ${\cal L} > 8\cdot 10^{30}$~cm$^{-2}$s$^{-1}$ for the centre of mass energy of 200 GeV and for 
${\cal L} >  2\cdot 10^{30}$~cm$^{-2}$s$^{-1}$ at $\sqrt{s} = 500$~GeV. Maximum rate of signal events of about 2~Hz and 10~Hz is 
expected for ${\cal L}  \sim  2\cdot 10^{32}$~cm$^{-2}$s$^{-1}$ at  $\sqrt{s} = 200$~GeV and $\sqrt{s} = 500$~GeV, respectively.
The signal to background rate is nearly 10 (100) for in whole region of considered instantaneous luminosities at centre of mass energy of 
200~GeV (500~GeV).

It is quite clear that larger signal to background ratio values could be obtained if the present Zero Degree Calorimeters were equipped
with an electromagnetic front part had a capability of both precise measurement of the electromagnetic energy and the
electromagnetic cascade position in the transverse plane. The former feature would allow decreasing the requirement on the
minimum photon energy and to improve its energy measurement and reconstruction while the latter would help to discriminate the 
events with production of neutral mesons decaying into multi-photon final state. 

One may conclude that the measurement of the cross-section for diffractive bremstrahlung at RHIC energies is feasible and, especially 
at the centre of mass energy of 500~GeV, for the instantaneous luminosity value of about $ 2\cdot 10^{30}$~cm$^{-2}$s$^{-1}$ the
event sample collected within a typical DAQ run would allow to obtain 1\% statistical precision. Such samples will help to experimentally 
determine values of the model parameters.

For the  instantaneous luminosity values greater than about $ 10^{30}$~cm$^{-2}$s$^{-1}$ the samples gathered with a typical run can 
be used for the RHIC instantaneous luminosity monitoring. Moreover, they will deliver a possibility to calibrate the relative energy loss 
measurement of a proton using the Roman Pot stations and to cross-check the stations alignment.

\section*{Acknowledgements}
We are very much indebted to P. Lebiedowicz and A. Szczurek for many stimulating discussions.
This work was supported in part by Polish National Science Centre grant UMO-2011/01/M/ST2/04126.


\begin{thebibliography}{999}
\bibitem{zeus} J. Andruszkow \textit{et al.}, Luminosity measurement in the ZEUS experiment, Acta. Phys. Pol. {\bf B32} (2001) 2025,\\
L. Adamczyk \textit{et al.}, Measurement of the luminosity in the ZEUS experiment at HERA II, Nucl. Instrm. Meth. {\bf A744} (2014) 80. 
\bibitem{khoze1}V. A. Khoze, J. W. L\"ams\"a, R. Orava, M. G. Ryskin, Forward Physics at the LHC: Detecting Elastic pp Scattering by Radiative Photons, JINST {\bf 6} (2011) P01005.
\bibitem{szczurek} P. Lebiedowicz and A. Szczurek, Exclusive diffractive photon bremsstrahlung at the LHC, Phys. Rev. {\bf D87} (2013) 114013.
\bibitem{dl} A. DOnnachie and P. V. Landshoff, Phys. Lett. {\bf B296} (1992) 227.
\bibitem{khoze2} V. A. Khoze, A. D. Martin, R. Orava, M. G. Ryskin. Luminosity measuring processes at the LHC, Eur. Phys. J {\bf C19} (2001) 313.
\bibitem{krasny1} M. W. Krasny, J. Chwastowski and K. S{\l}owikowski,   Luminosity measurement method for LHC: The theoretical precision and the experimental challenges, Nucl. Instrum. Meth.  {\bf A584} (2008) 42.
\bibitem{zdc} C. Adler \textit{et al.}, The RHIC Zero Degree Calorimeter, Nucl. Instrum. Meth. {\bf A470} (2001) 488,\\
M. B. Bitters \textit{et al.}, Analysis of STAR ZDC SMD Data for Polarimetry, unpublished.
\bibitem{rp} STAR Collaboration, L. Adamczyk \textit{et al.}, Single Spin Asymmetry AN in Polarized Proton-Proton Elastic Scattering at $\sqrt{s} = 200$~GeV , Phys.Lett. {\bf B719} (2013) 62.
\bibitem{kycia2} R. Kycia, J. Turnau, in preparation.
\bibitem{kycia1} R. Kycia, J. Turnau, R. Staszewski and J. Chwastowski, GenEx: A simple generator structure for exclusive processes in high energy collisions,  arXiv:{\bf 1411.6035 [hep-ph]}.
\bibitem{geant4} Geant4 Collab., S. Agostinelli et al., Nucl. Instrum. Meth. {\bf A506} (2003) 250,\\
Geant4 Collab., J. Allison et al., IEEE Trans. Nucl. Science {\bf 53} (2006) 270.
\bibitem{mad} F.~Schmidt, Mad-X User's Guide, CERN 2005, \\
BE/ABP Accelerator Physics Group: \verb+http://mad.web.cern.ch/mad/+.
\bibitem{sikora} R. Sikora, Master Thesis,AGH UST 2014, unpublished.
\bibitem{rafal} R. Staszewski and J. Chwastowski,  	
Transport Simulation and Diffractive Event Reconstruction at the LHC, Nucl. Instrum. Meth. {\bf A609} (2009) 136.
\bibitem{pythia} T. Sj\"ostrand \textit{et al.}, An Introduction to PYTHIA 8.2, arXiv:{\bf 1410.3012 [hep-ph]}, submitted to Computer Physics Communication,\\
T. Sj\"ostrand, S. Mrenna and P. Skands, PYTHIA 6.4 Physics and Manual, JHEP{\bf 05} )2006) 026.

\end{thebibliography}
\end{document}